\documentclass[a4paper]{article}

\usepackage{a4wide}

\usepackage[english]{babel} 
\usepackage{float}
\usepackage{amssymb,amsmath}
\usepackage{algorithm}
\usepackage{algpseudocode}
\usepackage{capt-of}
\usepackage{graphicx}

\usepackage{xspace}
\usepackage{tabularx}  

\renewenvironment{abstract}
 {\small
  \begin{center}
  \bfseries \abstractname\vspace{-.5em}\vspace{0pt}
  \end{center}
  \list{}{
    \setlength{\leftmargin}{.5cm}%
    \setlength{\rightmargin}{\leftmargin}%
  }%
  \item\relax}
 {\endlist}

\makeatletter
\renewcommand\fs@ruled{%
  \def\@fs@cfont{\rmfamily}%
  \let\@fs@capt\floatc@plain%
  \def\@fs@pre{\hrule height.8pt depth0pt \kern2pt}
  \def\@fs@post{\kern2pt\hrule\relax}
  \def\@fs@mid{\kern2pt\hrule\kern2pt}
  \def\@fs@post{}
  \let\@fs@iftopcapt\iffalse}
\makeatother

\usepackage{authblk}

\title{Automatic structure estimation of predictive models for symptom development}
\author[1]{Dennis Becker \thanks{dbecker@leuphana.de}}
\affil[1]{Department of Information Systems, Leuphana University}

\begin{document}

  \maketitle

\centerline{Short Paper Draft - arxiv.org }

\begin{abstract}

Online mental health treatment has the premise to meet the increasing demand for mental health treatment at a lower cost than traditional treatment.
However, online treatment suffers from high drop-out rates, which might negate their cost effectiveness.
Predictive models might aid in early identification of deviating clients which allows to target them directly to prevent drop-out and improve treatment outcomes.
We propose a two-staged multi-objective optimization process to automatically infer model structures based on ecological momentary assessment for prediction of future
symptom development. The proposed multi-objective optimization approach results in a temporal-causal network model with the best prediction performance for each concept. This allows for a selection of a disorder-specific model structure based on the envisioned field of application.
\end{abstract}

\section{Introduction}

Mental health disorders are an increasing problem in our society. In Europe, one in four people have reported that they suffered from at least one mental disorder during their lifetime~\cite{Alonso2004}
and it is estimated that in the USA approximately 26\% of the population suffers from at least one mental disorder~\cite{Kessler2005}.
These numbers are alarming and create increasing costs for our society~\cite{Lepine2011}.
The costs arise from the direct treatment but also due to the indirect costs of the loss of productivity or the workplace~\cite{Kessler2008}. The lost earnings in America are estimated to be \$193.2 billion per year~\cite{Insel2008}. The global costs of mental health were estimated at \$2.4 trillion, in 2010, and is predicted to increase to \$8.5 trillion by 2030~\cite{Bloom2011}.

As a result, online treatment gains an increasing importance for the delivery of mental health interventions~\cite{PMID:12358571,Cuijpers2008}. Online treatments utilize computerized cognitive behavioral therapy in form of brief therapy. These treatments are typically shorter than traditional therapy and specifically target a symptom or behavior.
Besides providing more people with the opportunity for treatment, online treatment promises to be more cost-effective than traditional face-to-face therapy~\cite{Tate2009}.
A therapist can supervise more clients in an online setting compared to a traditional face-to-face setting~\cite{DeGraaf2009,PROUDFOOT2003}, which lowers the intervention costs~\cite{Hedman2011}.
However, studies regarding the actual cost-effectiveness report mixed results \cite{VanBeugen2014,Donker2015}. One reason might be the considerably high drop-out rates in online treatment~\cite{VanBallegooijen2014}. A review study suggests that the drop-out rate of Internet-based treatment programs for psychological disorders ranges from 2 to 83\% and with a weighted average of 31\%~\cite{Melville2010}. An early identification of clients prone to drop-out might be crucial because dropped out clients are unlikely to recover and it negatively affects their attitude towards subsequent therapies \cite{White2010}.

Medical decision support systems could help in increasing the adherence and improving the outcome of online treatments~\cite{Vogenberg2009}.
In clinical psychology, the use of decision support systems has been evaluated and successfully used \cite{Trinanes2015}. The application of the evaluated systems ranges from automated screening using health records \cite{Rollman2001}, diagnosis support and treatment recommendations \cite{Thomas2004,Kurian2009}, to medication planning and scheduled screening reminders for physicians~\cite{Cannon2000}.
Online treatment has the advantage that a multitude of different data sources is already digitally available.
Several studies have shown that decision support systems can be useful for improving the quality of provided care, preventing errors, reducing finical costs, and saving human resources \cite{Kawamoto2005,Steyerberg2009}.
Integration of decision support systems into online treatment can provide similar advantages. Clients that are prone to deteriorate or do not respond to the treatment can receive an early intervention by the supervising therapist. The therapist could provide the deviating clients with additional or a different set of interventions that are more suited to their needs.
This would further strengthen the personal bond between the therapist and the client. An overall better treatment experience could also result in improved outcomes, which might reduce relapse risk and further treatment costs.


A regularly assessed measure that is suited for predictive model development is ecological momentary assessment (EMA)~\cite{Wichers2011}. EMA can consist of ratings of moods, thoughts, symptoms, or behavioral patterns and is conducted while clients are engaged in their typical daily routine~\cite{Wichers2011,Shiffman2008}. Typically, smart-phones are used to record these measures.
One particular type of model that allows to describe the interaction among EMA concepts is temporal-causal modelling~\cite{Treur2016}.
Such models capture the knowledge from psychological theories and describe the model as a set of differential equations that represent the interaction among the psychological concepts.
Based on this modeling approach, models for the emergence of depression due to stress factors or experience of extreme situations~\cite{Both2008}, the influence of different therapy strategies \cite{Both2011}, and social integration \cite{AltafHussainAbro2016} have been described.

Although these models provide the ability to capture psychological knowledge many choices are left to the person who designs them.
Furthermore, such models have many free parameters that allow them to provide a close fit to a variety of observed data. This makes them prone to overfitting the data which might negatively affect the prediction of future EMA development. In this paper, we propose to use a two-staged multi-objective genetic algorithm to explore the space of models based on the predictive performance. Multi-objective optimization provides the possibility to analyze the predictive performance of each individual concept. Because for different mental disorders the prediction performance of a particular concept might be more relevant than others. For example, in the case of depression the course of the concept mood might be the most reliable but for anxiety, a different concept could be more relevant.




In the following section, we discuss commonly assessed EMA measures, the design of temporal-causal network models, and a method for multi-objective optimization.
Following this, we present a description of our proposed two-stage optimization process for model structure estimation, as well as a plan for analyses and validation of the models estimated by our proposed approach. 
Finally, we illustrate our ideas for future research and list a number of potential limitations and difficulties with our current study.

\section{Research Setting and Approach}

In contrast to a model design that captures psychological knowledge and explains observed data, we aim to explore a model design that predicts future concept development.
In this section, we describe typically assessed EMA data and the methods that we intend to use to explore predictive models in the application of future symptom development.

\subsection{Ecological Momentary Assessment}
\label{sec:emadata}

EMA is part of a clinical and online treatment of mental disorders. Especially, in online treatment that is delivered using a smart-phone, it is easily implemented and inquired.
EMA concepts are inquired by presenting the client a series of questions that are rated by the client on a numeric scale from 1 to 10.
An overview of commonly assessed EMA concepts and the inquiry question is shown in Table~\ref{tab:ema}.

\begin{table}[H]
\begin{center}
\caption{Example of commonly used EMA concepts}
\label{tab:ema}
\begin{tabular}{l|l}
\hline
\textbf{EMA concept} & \textbf{Assessment question} \\
\hline
Mood               &       How is your mood right now?  \\
Worry              &       How much do you worry about things at the moment? \\
Self-Esteem        &       How good do you feel about yourself right now? \\
Sleep              &       How did you sleep tonight?  \\
Activities done    &       To what extent have you carried out enjoyable activities today?  \\
Enjoyed activities &       How much have you enjoyed the days activities? \\ 
Social contact     &       How much have you been involved in social interactions today?\\
\hline
\end{tabular}
\end{center}

\end{table}

The amount of assessed EMA concepts and the number of inquiries varies among studies. One reason is that assessment can be done multiple times throughout the day for example in the morning and the evening, or only once each day. A reason to reduce the number of EMA inquirements is because a too high workload of EMA assessment might be considered as intrusive and clients might become reluctant to enter their EMA measures. 
Therefore, some online interventions do not assess all the concepts every day.
Another point to consider is that in contrast to clients that use a standalone PC to process the interventions, smart-phones can provide a reminding function. This increases the likelihood of clients responding to these questions. Additionally, online treatments vary in length of active treatment, which can range from 4 to 15 weeks.
This shows that there might be variability in the granularity and length of the available EMA data, also with respect to clients that drop-out early. Therefore, it is expected that the amount of available data varies among clients and contains many missing values.

\subsection{Temporal-causal Network Models}

Temporal-causal network models provide the possibility to represent biological, physical, or social networks, which inherently contain cycles.
These models allow to capture the available knowledge in form of a network representation and provide simulations for each represented concept.
The individual concepts, or nodes in this network, are connected by differential equations that describe the change of the concept in relation to the connected concepts.
For example, the change of the concept Y can be described as the following equation:
\begin{equation}
    \frac{dY(t)}{dt} = \eta_Y [C_Y(\omega_{X_1,Y}X_1(t),..., \omega_{X_k,Y}X_k(t)) - Y(t)].
\end{equation}

Where the rate of change is defined by three components:
\begin{itemize}
    \item A weight $\omega$ that defines the strength of the causal relation to the concept.
    \item The factor $\eta$ that defines the speed of change of the concept based on the causal impact.
    \item A combining function $C(\cdot)$ that combines the causal impacts of connected concepts.
\end{itemize}

\begin{figure}[H]
	\centering
  \includegraphics[width=0.5\textwidth]{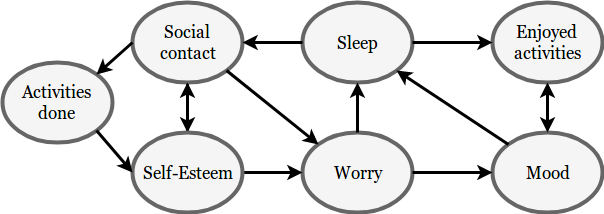}
	\caption{Possible network model for EMA interaction}
	\label{fig:model}
\end{figure}

For the choice of combining function, a variety of standard combining functions exist. These standard combining functions are building blocks and can be used during the design of a model. These functions are based, for example, on the sum, product, max, min or a simple logistic function. However, some of these standard combining functions such as the simple logistic function require additional parameters.
An example network representation for the EMA data is illustrated in Figure~\ref{fig:model}. 

Although humans tend to understand a network representation more intuitively, for a computer this representation is more difficult.
But the model can also easily be represented as a matrix, as shown in Table~\ref{tab:model}, which makes it easier to automatically generate different models.

\begin{table}[H]
\centering
\caption{Matrix representation of the model shown in Figure~\ref{fig:model} } 
\label{tab:model}
\begin{tabular}{r|cccccccc}
  \hline
    to       & Mood  & Worry& Sleep & Self-Esteem &Social  & Enjoyed  & Activities  \\ 
    from   \quad \quad    &   & &  &  & contact &  activities &  done \\ 
  \hline
  Mood     &    & & $\omega_{m,s}$ &  &  &$\omega_{m,ea}$  &   \\ 
  Worry       & $\omega_{w,m}$     & & $\omega_{w,s}$ &  &  &  &         \\ 
  Sleep        &       & &  &  &$\omega_{s,sc}$ & $\omega_{s,ea}$    &    \\ 
  Self-Esteem  &   & $\omega_{se,w}$&  &  &$\omega_{se,sc}$ &   &\\ 
  Social contact  &    & $\omega_{sc,w}$&  & $\omega_{sc,se}$ &  &   &$\omega_{sc,ad}$        \\ 
  Enjoyed activities & $\omega_{ea,m}$  & &  & & &  &       \\ 
 Activities done &    & &  & $\omega_{ad,se}$ &  & & \\ 
\hline
  $\eta_Y$  & $\eta_m$   & $\eta_w$& $\eta_s$ & $\eta_{se}$ & $\eta_{sc}$ & $\eta_{ea}$&$\eta_{ad}$          \\ 
  $C_Y(\cdot)$ & $C_{m}(\cdot)$   & $C_{w}(\cdot)$& $C_{s}(\cdot)$ & $C_{se}(\cdot)$ &$C_{sc}(\cdot)$ & $C_{ea}(\cdot)$ & $C_{ad}(\cdot)$          \\ 
  \hline
\end{tabular}
\end{table}

The matrix representation of the model shows clearly what parameters need to be estimated, and one can always switch between both representations. The network representation allows researchers to intuitively evaluate the created model, whereas the matrix representation is more suited for computational purposes.

\subsection{Multi-objective Optimisation}
\label{sec:nsga2}

For the two staged model structure optimization, we utilize the NSGA2 (nondominated sorting genetic algorithm II)~\cite{Deb2002} optimization algorithm. It will be used for the estimation of the model structure and client specific model parameters. NSGA2 is a genetic algorithm for multi-objective optimization that estimates the parameters with respect to multiple objectives and can be used for any type of optimization. Evolutionary algorithms, in particular, have been proven to provide acceptable solutions and are well suited for discontinuous and multi-modal optimization problems~\cite{Goldberg1991,CoelloCoello1999}. They have already been applied to a variety of real-world optimization tasks, such as design optimization \cite{Konig1999,Lee1996}, circuit design~\cite{Miller1997,Soleimani2011}, and routing \cite{Knowles2000}.

An evolutionary algorithm is a population-based search and optimization method that mimics the process of natural evolution~\cite{holland1975}.
Such algorithms start with an initial population, where each individual of the population is a solution to the problem. Generations are created successively based on the best solutions so far.
In each generation of the algorithm, the parameters of the best individuals are merged to generate new solutions. This merging process of parameters is called crossover.
The quality of these solutions is then estimated by evaluating the prediction error. The solutions with a higher prediction error are replaced by new solutions that provide a lower prediction error. To create additional diversity in this mechanism, mutation is introduced. Mutation applies small random changes to the parameters of a newly created solution during the crossover process. 
Typically, the search behavior of the algorithm can be summarized as exploration and exploitation. During exploration, large changes in the search space allow identifying new regions with a lower error on the cost function. Over the course of generations, the diversity of the population will be reduced and smaller changes allow to refine the already estimated solutions to further reduce the error.

However, NSGA2 has a one-dimensional representation of the parameters that are to be optimized. This representation might not perform well for our two-dimensional model matrix. Models that are similar might not be represented in terms of continuity of the parameter space. This could negatively affect the model structure search. Therefore, we consider integrating a two-dimensional parameter representation into NSGA2. This two-dimensional parameter encoding representation requires to replace the crossover and mutation operations and could provide more efficient exploration and optimization in the model space~\cite{Tsai2015}.

\section{Model Estimation and Validation}

In order to estimate a model that predicts future data well, we first have to define what time granularity is to be predicted.
Since study data is likely to provide daily measures, we aim at the prediction of a concept value for each day. If multiple measures are available for each day, the average value is assumed as the measure.
Next, we have to define for what time frame the predictions are supposed to be accurate. 

For the evaluation of the model performance, the one-step-ahead prediction is often considered. This means that the concepts for the next day are predicted and the error is estimated accordingly. Then the measures of this day are used to adapt the model, and a prediction for the next day is given. However, for an online setting, the prediction of the following day might be too short. Typically, a supervising therapist could not check on each client every day. 
To ensure cost-effectiveness and reduce the effort in supervising many clients, it would be more likely that a therapist checks each or every second week on a client. In addition, it might be unlikely that a deterioration or drop-out happens overnight. It is more likely that a client's attitude towards the treatment changes over the course of the interventions. 
Some online interventions advise the clients to complete at least one but preferably two interventions per week. This would make a prediction of a two-week time frame more reasonable to gauge the clients overall development. The length of the training time series required for learning of the model structure might depend on the number of available measures in each week. We assume that at least 3 weeks of data are necessary to learn the underlying process, where a longer time series might be beneficial.

To learn a model structure that is suited to predict future concept development, we separate the data into a training, test, and validation set. This separation allows to fit the model towards the test set, but at the same time remains some data for actual testing of the predictive performance of unseen data that has not been used for learning of the model structure.
A model that accurately predicts the next week does not necessarily have to be the best model to predict the following two weeks in advance.
It is also questionable if a model that predicts the second week of treatment is as reliable for the prediction after 10 weeks into the treatment. With progression in the treatment, the client might experience a reduction in symptoms and change in concepts might stabilize. 
However,  we are focussing on the model structure because model parameters are likely to vary among clients and also over time.
Therefore, based on the model structure individual client parameters need to be estimated.

\begin{figure}[H]
	\centering
  \includegraphics[width=0.75\textwidth]{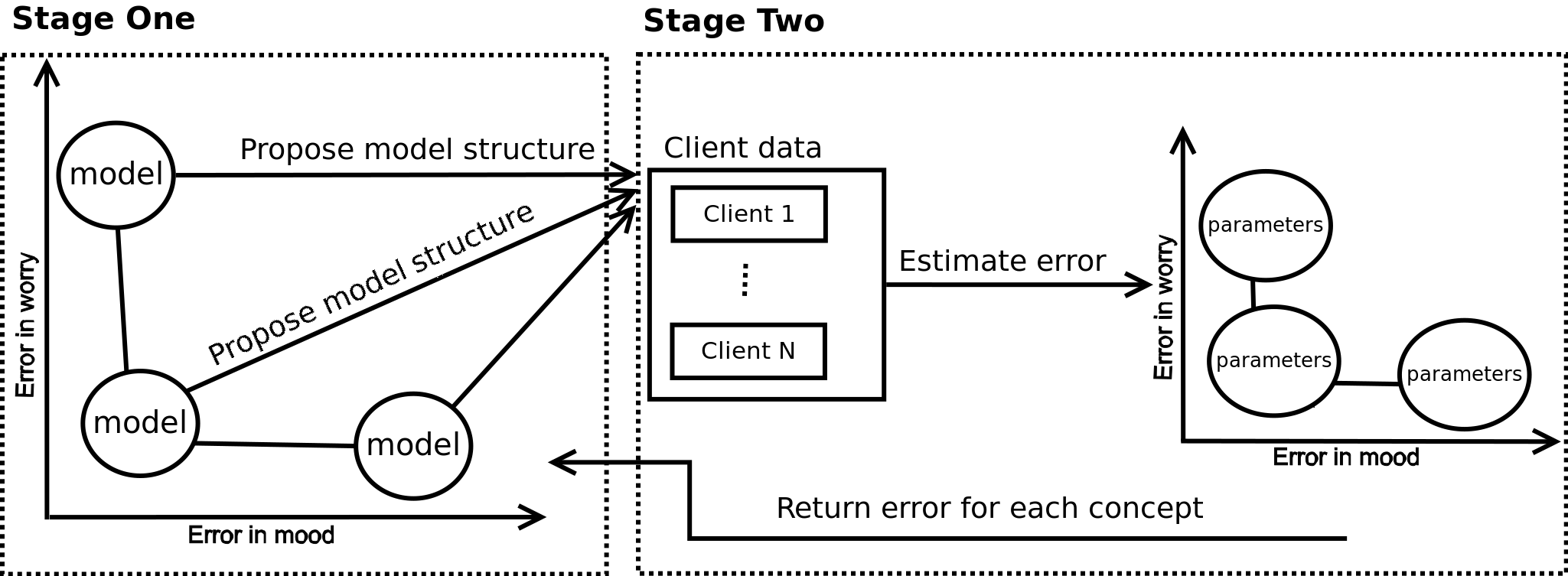}
	\caption{Illustration of the two-staged model structure estimation approach}
	\label{fig:approach}
\end{figure}

This suggests a two-stage approach, which is illustrated in Figure~\ref{fig:approach}.
Both stages utilize the NSGA2 multi-objective optimization algorithm, where the first stage estimates a model structure and the second stage client individual model parameters.
In the first stage, the NSGA2 optimization algorithm is used to propose a variety of model structures. 
For these model structures, no particular weights are required, rather only the connections and combination functions are proposed in the first stage.
Each model structure has a specific error for each concept. In the illustration of the approach, we only show two dimensions instead of all seven.
The error of every concept has to be estimated for each client, which requires individual parameters for each client. These client-specific model parameters account for potentially time-dependent influences and lead to the second stage.
In the second stage, parameters for the models are estimated for each client and the prediction error for each concept is estimated. 
The parameters are estimated from the training set, and the prediction error is estimated on the test set.
Each parameter setting provides a different error for each concept. Therefore, we take the minimal error of each concept. These errors are from the same model but are results of different parameter settings, as indicated in the illustration of the approach. This leads to an estimation of an error for each client and each concept. The errors for each concept are summed up and reported back to the first stage. With these errors, the first stage algorithm can then continue the search for a model structure that reduces the overall prediction error on the test set.

We aim to learn a model structure that captures the dynamics of the data. This procedure might be beneficial because it aims at learning the underlying task for each concept and not a model structure for each client. An individual model structure for each client might also be misleading since the data quality among clients varies and contains noise.
However, this process will not result in a single model, rather than in the number of the chosen population size of the first stage NSGA2 algorithm. To evaluate the resulting models, we consider the models that have the lowest prediction error in each concept. This results in 7 estimated models, where each model provides the lowest prediction error in one concept. The network representation of these models allows to compare them to existing knowledge about the theoretical interaction among these concepts and comparison of the structure of the resulting models might provide insight into the structural composition of these models. Furthermore, we evaluate the resulting prediction performance of these models on the validation set and compare the prediction performance against reference measures such as a regression model, dynamic linear models, and mean value prediction.

\section{Concluding Remarks} 

In this short paper, we propose an approach for automated predictive model estimation based on EMA data. 
Specifically, we propose to apply multi-parameter optimization, which results in a model for each concept that provides the best prediction performance.
This would allow to chose a model based on the intended purpose. For the case of depression, the mood development might be the most important whereas for social anxiety a prediction of social contact might be a more reliable indicator. 
By utilizing temporal-causal network models, the resulting models consist of a network structure that captures the psychological knowledge and underlying process.
In contrast to other machine learning approaches such as neuronal networks or support vector machines, this makes it possible for humans to understand the resulting model and check the connections against existing results from psychological research. This allows to validate if the resulting models coincide with results from the psychological literature.

Regarding predictive model development for use in online treatment, estimation of a model structure is only the first step.
After deriving promising model structures, the next step could be to research how these tools can be integrated into an online treatment setting, such that a supervising therapist can use them beneficially on a client individual level. Furthermore, a prediction of low concepts or rapid changes in the concepts have to be detected automatically, and it has to be analyzed how to derive treatment decisions from these predictions.



  
This research-in-progress study has a number of limitations.
First, the proposed approach might be computational very demanding especially with an increasing number of clients. Because for each model that is proposed by the first stage genetic algorithm, the predictive performance for all clients needs to be estimated. This number would further increase if we assume that a client's time series can be split into multiple chunks of the targeted two weeks time frame.
To overcome this problem, one could consider an alternative route by estimating the model structure directly in one step. This approach would be straightforward, however, to avoid fully connected models, one has to add an additional complexity term that enforces sparse models. This complexity term might consist of the sum of the connection weights.
Second, the amount of available EMA data depends on the clients' compliance to the assessment protocol and treatment engagement. Therefore, the measurements for concepts might be sparse, which could complicate the model structure estimation.
Finally, the prediction performance of the estimated models has to be superior to than the reference measures and provide reliable predictions. A model that does not provide reliable predictions about the future might provide insides into the interaction of EMA concepts but not be beneficial for deriving treatment decisions.



\newpage

\bibliographystyle{plain}

\bibliography{references.bib}

\begin{thebibliography}{10}

\bibitem{Alonso2004}
J~Alonso, M~C Angermeyer, S~Bernert, R~Bruffaerts, T~S Brugha, H~Bryson,
  G~de~Girolamo, R~Graaf, K~Demyttenaere, I~Gasquet, J~M Haro, S~J Katz, R~C
  Kessler, V~Kovess, J~P L{\'{e}}pine, J~Ormel, G~Polidori, L~J Russo,
  G~Vilagut, J~Almansa, S~Arbabzadeh-Bouchez, J~Autonell, M~Bernal, M~a
  Buist-Bouwman, M~Codony, a~Domingo-Salvany, M~Ferrer, S~S Joo,
  M~Mart{\'{i}}nez-Alonso, H~Matschinger, F~Mazzi, Z~Morgan, P~Morosini,
  C~Palac{\'{i}}n, B~Romera, N~Taub, and W~a~M Vollebergh.
\newblock {Prevalence of mental disorders in Europe: results from the European
  Study of the Epidemiology of Mental Disorders (ESEMeD) project.}
\newblock {\em Acta psychiatrica Scandinavica. Supplementum}, 109(420):21--27,
  2004.

\bibitem{AltafHussainAbro2016}
Michel~Klein {Altaf Hussain Abro}.
\newblock {Validation of a Computational Model for Mood and Social
  Integration}.
\newblock {\em Lecture Notes in Computer Science}, 10046(November
  2016):361--375, 2016.

\bibitem{Bloom2011}
David~E. Bloom, Elizabeth Cafiero, Eva Jan{\'{e}}-Llopis, Shafika
  Abrahams-Gessel, Lakshmi {Reddy Bloom}, Sana Fathima, Andrea {B. Feigl}, Tom
  Gaziano, Ali Hamandi, Mona Mowafi, Danny O'Farrell, Emre Ozaltin, Ankur
  Pandya, Klaus Prettner, Larry Rosenberg, Benjamin Seligman, Adam~Z. Stein,
  Cara Weinstein, and Jonathan Weiss.
\newblock {The Global Economic Burden of Noncommunicable Diseases}.
\newblock {\em World Economic Forum}, (September):1--46, 2011.

\bibitem{Both2008}
Fiemke Both, Mark Hoogendoor, Michel Klein, and Jan Treur.
\newblock {Modeling the Dynamics of Mood and Depression}.
\newblock In {M. Ghallab}, {C.D. Spyropoulos}, {N. Fakotakis}, and {N.
  Avouris}, editors, {\em Proceedings of the 18th European Conference on
  Artificial Intelligence, ECAI'08}, pages 266--270, 2008.

\bibitem{Both2011}
Fiemke Both and Mark Hoogendoorn.
\newblock {Utilization of a virtual patient model to enable tailored therapy
  for depressed patients}.
\newblock {\em Neural Information Processing}, pages 700--710, 2011.

\bibitem{Cannon2000}
D.~S. Cannon and S.~N. Allen.
\newblock {A Comparison of the Effects of Computer and Manual Reminders on
  Compliance with a Mental Health Clinical Practice Guideline}.
\newblock {\em Journal of the American Medical Informatics Association},
  7(2):196--203, mar 2000.

\bibitem{PMID:12358571}
Helen Christensen and Kathleen~M Griffiths.
\newblock The prevention of depression using the internet.
\newblock {\em The Medical journal of Australia}, 177 Suppl:S122—5, October
  2002.

\bibitem{CoelloCoello1999}
Carlos~A {Coello Coello}.
\newblock {An Updated Survey of Evolutionary Multiobjective Optimization
  Techniques : State of the Art and Future Trends}.
\newblock {\em 1999 Congress on Evolutionary Computation}, 1(4):3--13, 1999.

\bibitem{Cuijpers2008}
Pim Cuijpers, Straten~A Van, Gerhard Andersson, Annemieke van Straten, Gerhard
  Andersson, Straten~A Van, and Gerhard Andersson.
\newblock {Internet-administered cognitive behavior therapy for health
  problems: a systematic review}.
\newblock {\em J Behav.Med}, 31(0160-7715 (Print)):169--177, 2008.

\bibitem{DeGraaf2009}
L.~E. {De Graaf}, S~a~H Gerhards, A~Arntz, H~Riper, J~F~M Metsemakers, S~M a~a
  Evers, J~L Severens, G~Widdershoven, and M~J~H Huibers.
\newblock {Clinical effectiveness of online computerised cognitive-behavioural
  therapy without support for depression in primary care: Randomised trial}.
\newblock {\em British Journal of Psychiatry}, 195(1):73--80, jul 2009.

\bibitem{Deb2002}
Kalyanmoy Deb, Amrit Pratap, Sameer Agarwal, and T.~Meyarivan.
\newblock {A fast and elitist multiobjective genetic algorithm: NSGA-II}.
\newblock {\em IEEE Transactions on Evolutionary Computation}, 6(2):182--197,
  2002.

\bibitem{Donker2015}
T.~Donker, M.~Blankers, E.~Hedman, B.~Lj{\'{o}}tsson, K.~Petrie, and
  H.~Christensen.
\newblock {Economic evaluations of Internet interventions for mental health: a
  systematic review}.
\newblock {\em Psychological Medicine}, (AUGUST):1--20, 2015.

\bibitem{Goldberg1991}
David~E Goldberg, Kalyanmoy Deb, and James~H Clark.
\newblock {Genetic Algorithms, Noise, and the Sizing of Populations}.
\newblock {\em Complex Systems}, 6:333--362, 1991.

\bibitem{Hedman2011}
Erik Hedman, Erik Andersson, Brj{\'{a}}nn Lj{\'{o}}tsson, Gerhard Andersson,
  Christian R{\"{u}}ck, and Nils Lindefors.
\newblock {Cost-effectiveness of Internet-based cognitive behavior therapy vs.
  cognitive behavioral group therapy for social anxiety disorder: Results from
  a randomized controlled trial}.
\newblock {\em Behaviour Research and Therapy}, 49(11):729--736, 2011.

\bibitem{holland1975}
John~H. Holland.
\newblock {\em {Adaptation in Natural and Artificial Systems}}.
\newblock University of Michigan Press, Ann Arbor, MI, USA, 1975.

\bibitem{Insel2008}
Tr~Insel.
\newblock {Assessing the economic costs of serious mental illness}.
\newblock {\em Am. J. Psychiat.}, 165(6):663--665, 2008.

\bibitem{Kawamoto2005}
K.~Kawamoto, C.~A. Houlihan, E.~A. Balas, and D.~F. Lobach.
\newblock {Improving clinical practice using clinical decision support systems:
  a systematic review of trials to identify features critical to success}.
\newblock {\em BMJ (Clinical research ed.)}, 330(7494):765, 2005.

\bibitem{Kessler2005}
Rc~Kessler and Wt~Chiu.
\newblock {Prevalence, Severity, and Comorbidity of Twelve-month DSM-IV
  Disorders in the National Comorbidity Survey Replication (NCS- R)}.
\newblock {\em Archives of general psychiatry}, 62(6):617--627, 2005.

\bibitem{Kessler2008}
Ronald~C Kessler, Steven Heeringa, Matthew~D Lakoma, Maria Petukhova, Agnes~E
  Rupp, Michael Schoenbaum, Philip~S Wang, and Alan~M Zaslavsky.
\newblock {Individual and societal effects of mental disorders on earnings in
  the United States: results from the national comorbidity survey replication.}
\newblock {\em The American journal of psychiatry}, 165(6):703--11, 2008.

\bibitem{Knowles2000}
J~D Knowles and D~W Corne.
\newblock {Heuristics for Evolutionary Off-line Routing in Telecommunications
  Networks}.
\newblock In {\em Proceedings of the Genetic and Evolutionary Computation
  Conference}, number January, pages 574--581, 2000.

\bibitem{Konig1999}
Oliver K{\"{o}}nig and G~Fadel.
\newblock {Application of genetic algorithms in the design of multi-material
  structures manufactured in rapid prototyping}.
\newblock In {\em Solid Freeform Fabrication Symposium}, pages 209--218, 1999.

\bibitem{Kurian2009}
Benji~T Kurian, Madhukar~H Trivedi, Bruce~D Grannemann, Cynthia~a Claassen,
  Ella~J Daly, and Prabha Sunderajan.
\newblock {A computerized decision support system for depression in primary
  care.}
\newblock {\em Primary care companion to the Journal of clinical psychiatry},
  11(4):140--146, 2009.

\bibitem{Lee1996}
Jongsoo Lee and Prabhat Hajela.
\newblock {Parallel genetic algorithm implementation in multidisciplinary rotor
  blade design}.
\newblock {\em Journal of Aircraft}, 33(5):962--969, 1996.

\bibitem{Lepine2011}
Jean~Pierre L{\'{e}}pine and Mike Briley.
\newblock {The increasing burden of depression}.
\newblock {\em Neuropsychiatric Disease and Treatment}, 7(SUPPL.):3--7, 2011.

\bibitem{Melville2010}
Katherine~M Melville, Leanne~M Casey, and David~J Kavanagh.
\newblock {Dropout from Internet-based treatment for psychological disorders.}
\newblock {\em The British journal of clinical psychology / the British
  Psychological Society}, 49(Pt 4):455--71, nov 2010.

\bibitem{Miller1997}
JF~Miller, P~Thomson, and T~Fogarty.
\newblock {Designing electronic circuits using evolutionary algorithms.
  arithmetic circuits: A case study}.
\newblock {\em Genetic algorithms and evolution strategies in engineering and
  computer science}, 8, 1997.

\bibitem{PROUDFOOT2003}
J.~Proudfoot, D.~Goldberg, A.~Mann, B.~Everitt, I.~Marks, and J.~A. Gray.
\newblock {Computerized, interactive, multimedia cognitive-behavioural program
  for anxiety and depression in general practice}.
\newblock {\em Psychological Medicine}, 33(2):S0033291702007225, 2003.

\bibitem{Rollman2001}
Bruce~L. Rollman, Barbara~H. Hanusa, Trae Gilbert, Henry~J. Lowe, Wishwa~N.
  Kapoor, and Herbert~C. Schulberg.
\newblock {The Electronic Medical Record}.
\newblock {\em Archives of Internal Medicine}, 161(2):189, 2001.

\bibitem{Shiffman2008}
Saul Shiffman, Arthur~A. Stone, and Michael~R. Hufford.
\newblock {Ecological momentary assessment}.
\newblock {\em Annual review of clinical psychology}, 4(5):1--32, 2008.

\bibitem{Soleimani2011}
Parisa Soleimani, Reza Sabbaghi-Nadooshan, Sattar Mirzakuchaki, and Mahdi
  Bagheri.
\newblock {Using Genetic Algorithm in the Evolutionary Design of Sequential
  Logic Circuits}.
\newblock {\em IJCSI International Journal of Computer Science Issues},
  8(3):1--7, 2011.

\bibitem{Steyerberg2009}
E.W. Steyerberg.
\newblock {Applications of prediction models}.
\newblock In {\em Springer}, volume~36, pages 11--31. 2009.

\bibitem{Tate2009}
Deborah~F. Tate, Eric~A. Finkelstein, Olga Khavjou, and Alison Gustafson.
\newblock {Cost effectiveness of internet interventions: Review and
  recommendations}.
\newblock {\em Annals of Behavioral Medicine}, 38(1):40--45, 2009.

\bibitem{Thomas2004}
H~V Thomas, G~Lewis, M~Watson, T~Bell, I~Lyons, K~Lloyd, S~Weich, and D~Sharp.
\newblock {Computerised patient-specific guidelines for management of common
  mental disorders in primary care: A randomised controlled trial}.
\newblock {\em British Journal of General Practice}, 54(508):832--837, 2004.

\bibitem{Treur2016}
Jan Treur.
\newblock {Dynamic modeling based on a temporal-causal network modeling
  approach}.
\newblock {\em Biologically Inspired Cognitive Architectures},
  16(April):131--168, 2016.

\bibitem{Trinanes2015}
Yolanda Tri{\~{n}}anes, Gerardo Atienza, Arturo Louro-Gonz{\'{a}}lez, Elena
  De-las Heras-Li{\~{n}}ero, Mar{\'{i}}a Alvarez-Ariza, and Diego~J. Palao.
\newblock {Development and impact of computerised decision support systems for
  clinical management of depression: A systematic review}.
\newblock {\em Revista de Psiquiatr{\'{i}}a y Salud Mental (English Edition)},
  8(3):157--166, 2015.

\bibitem{Tsai2015}
Ming~Wen Tsai, Tzung~Pei Hong, and Woo~Tsong Lin.
\newblock {A two-dimensional genetic algorithm and its application to aircraft
  scheduling problem}.
\newblock {\em Mathematical Problems in Engineering}, 2015, 2015.

\bibitem{VanBallegooijen2014}
Wouter {Van Ballegooijen}, Pim Cuijpers, Annemieke {Van Straten}, Eirini
  Karyotaki, Gerhard Andersson, Jan~H. Smit, and Heleen Riper.
\newblock {Adherence to internet-based and face-to-face cognitive behavioural
  therapy for depression: A meta-analysis}.
\newblock {\em PLoS ONE}, 9(7), 2014.

\bibitem{VanBeugen2014}
Sylvia {Van Beugen}, Maaike Ferwerda, Dane Hoeve, Maroeska~M. Rovers, Saskia
  {Spillekom-Van Koulil}, Henri??t {Van Middendorp}, and Andrea W~M Evers.
\newblock {Internet-based cognitive behavioral therapy for patients with
  chronic somatic conditions: A meta-analytic review}.
\newblock {\em Journal of Medical Internet Research}, 16(3):1--15, 2014.

\bibitem{Vogenberg2009}
F~Randy Vogenberg.
\newblock {Predictive and Prognostic Models: Implications for Healthcare
  Decision-Making in a Modern Recession}.
\newblock {\em Am Health Drug Benefits}, 2(6), 2009.

\bibitem{White2010}
Kamila~S White, Laura~B Allen, David~H Barlow, Jack~M Gorman, M~Katherine
  Shear, and Scott~W Woods.
\newblock {Attrition in a multicenter clinical trial for panic disorder.}
\newblock {\em The Journal of nervous and mental disease}, 198(9):665--671,
  2010.

\bibitem{Wichers2011}
M.~Wichers, C.~J~P Simons, I.~M~A Kramer, J.~A. Hartmann, C.~Lothmann,
  I.~Myin-Germeys, A.~L. van Bemmel, F.~Peeters, Ph~Delespaul, and J.~van Os.
\newblock {Momentary assessment technology as a tool to help patients with
  depression help themselves}.
\newblock {\em Acta Psychiatrica Scandinavica}, 124(4):262--272, 2011.

\end{thebibliography}

\end{document}